\documentstyle[aps,prl,epsf,twocolumn]{revtex}
\begin{document}
\twocolumn[\hsize\textwidth\columnwidth\hsize\csname@twocolumnfalse%
\endcsname] 
\def\bea{\begin{eqnarray}}
\def\eea{\end{eqnarray}}
\def\d{\delta}
\def\nn{\nonumber}
\def\r{\rho}
\def\la{\langle}
\def\ra{\rangle}
\def\e{\epsilon}
\def\n{\eta}

{\bf {\noindent Comment on ``Can Disorder Induce a Finite
Thermal Conductivity in $1D$ Lattices?''}}

\narrowtext

In a recent letter, Li et al \cite{li} have reported that the steady state of
the disordered harmonic chain is not unique and depends on initial
conditions. Their claim is based on a molecular dynamics simulation
using Nose-Hoover thermostats to model the heat baths.   
We point out that the uniqueness of the nonequilibrium
steady state of the disordered harmonic chain for a large class of
heat baths has been proven exactly \cite{lebo,spohn}. In this note we consider
a particular case where it is easy to explicitly demonstrate the
uniqueness of the steady state.  
This is the case where the heat bath is modelled by Langevin dynamics. 
The results of Li et al appear to be either due to 
insufficient equilibration times  or an artifact of using Nose-Hoover
thermostats.

Consider heat conduction through a one-dimensional disordered
harmonic chain. Particles $i=1,2...N$ with random masses 
are connected by harmonic springs of equal strengths. 
The Hamiltonian of the system is thus
\bea
H=\sum_{l=1}^N \frac{p_l^2}{2 m_l} + \sum_{l=0}^{N} \frac{(x_l-x_{l+1})^2}{2}  
\eea 
where $\{x_l\}$ are the displacements of the particles about their
equilibrium positions, $\{p_l\}$ their momenta and  $\{m_l\}$
are the random masses. We put the boundary conditions $x_0=x_{N+1}=0$.
The particles in the bulk evolve through the classical equations of
motion while the boundary particles, namely particles $1$ and $N$ are
coupled to langevin heat baths as in Ref. \cite{lebo}. 

Thus we consider the following equations of motion for the particles:
\bea
m_1 \ddot{x_1} &=&- 2 x_1+x_2- \dot{x}_1+\n_L(t) \nn \\
m_l \ddot{x_l} &=& - (2 x_l-x_{l-1}-x_{l+1})~~~~~l=2,3...(N-1)  \nn \\
m_N \ddot{x_N} &=& - 2 x_N+x_{N-1})-\dot{x}_N+\n_R(t),  
\label{eqmot}
\eea
where $\n_L$ and $\n_R$ are gaussian white noises with the
correlations $\la \n_{L,R}(t) \n_{L,R}(t')\ra= 2 T_{L,R}
\delta(t-t')$. 
Denoting the coordinates and momenta collectively by
the variables $q_l$ so that $\{ q_1,q_2...q_{2N} \}= \{
x_1,x_2...x_N,p_1,p_2...p_N \}$ Eqs.~(\ref{eqmot}) can be written in the form:
\bea
\dot{q}_l=-\sum_{k=1}^{2N} a_{lk}q_k+\n_l 
\label{neqmot}
\eea
where the vector $\n$ has all elements zero except $\n_{N+1}=\n_L$ and
$\n_{2 N}=\n_R$ and the $2N \times 2N$ matrix $a$ is given by:
\bea
&& a= \left( \begin{array}{cc}
0 & J \\ \Phi & \gamma \end{array} \right) ~~~~{\rm{with}} \\
&& J_{kl}= -\delta_{k,l}/m_k;~\Phi_{kl}=2 \delta_{k,l}-\delta_{k
,l-1}-\delta_{k,l+1} \nn \\
&& \gamma_{kl}=\delta_{k,l}( \delta_{k,1}/m_1+\delta_{k,N} /m_N ). \nn
\eea
In the steady state $\la d(q_kq_l)/dt \ra =0$. From this and using 
Eq.~(\ref{neqmot}) we get the matrix equation
\bea
a.b+b.a^T=d,
\label{mateq}
\eea
where $b$ is the correlation matrix with elements $b_{kl}=\la q_k q_l
\ra$ and $d_{kl}=\delta_{k,l} (2 T_L \delta_{k,N+1}+2 T_R \delta_{k,
2N})$. We can invert this equation to obtain $b$ and thus all the
moments including the local temperatures $T_l=p_l^2/m_l$.
The uniqueness of the steady state then
depends on whether or not Eq.~(\ref{mateq}) has a unique inverse. For
chains of finite length and for given disorder realizations it is 
easy to verify numerically that Eq.~(\ref{mateq}) does have a unique
inverse. We also find that a molecular dynamics 
simulation (using a simple Euler discretization), with langevin heat
baths, reproduces  the exact temperature profile and is independent of
initial conditions. This is shown in Fig.~(\ref{tprof}). For the $N=20$
lattice, averaging over $10^7$ time units is sufficient to achieve
steady-state values. The $N=40$ data is averaged over $10^8$ time
units. We find that equilibration times increase rapidly with system
size. 

Finally we note that for Langevin heat bath dynamics, the unique steady
state distribution is infact known exactly \cite{lebo} and is a
gaussian given by: 
\bea
P(\{ q_l \} ) =(2 \pi)^{-N} Det[b]^{-1/2} e^{-\frac{1}{2}\sum
b^{-1}_{lm}q_l q_m}.
\eea

\vbox{
%\vspace{0.5cm}
\epsfxsize=8.0cm
\epsfysize=6.0cm
\epsffile{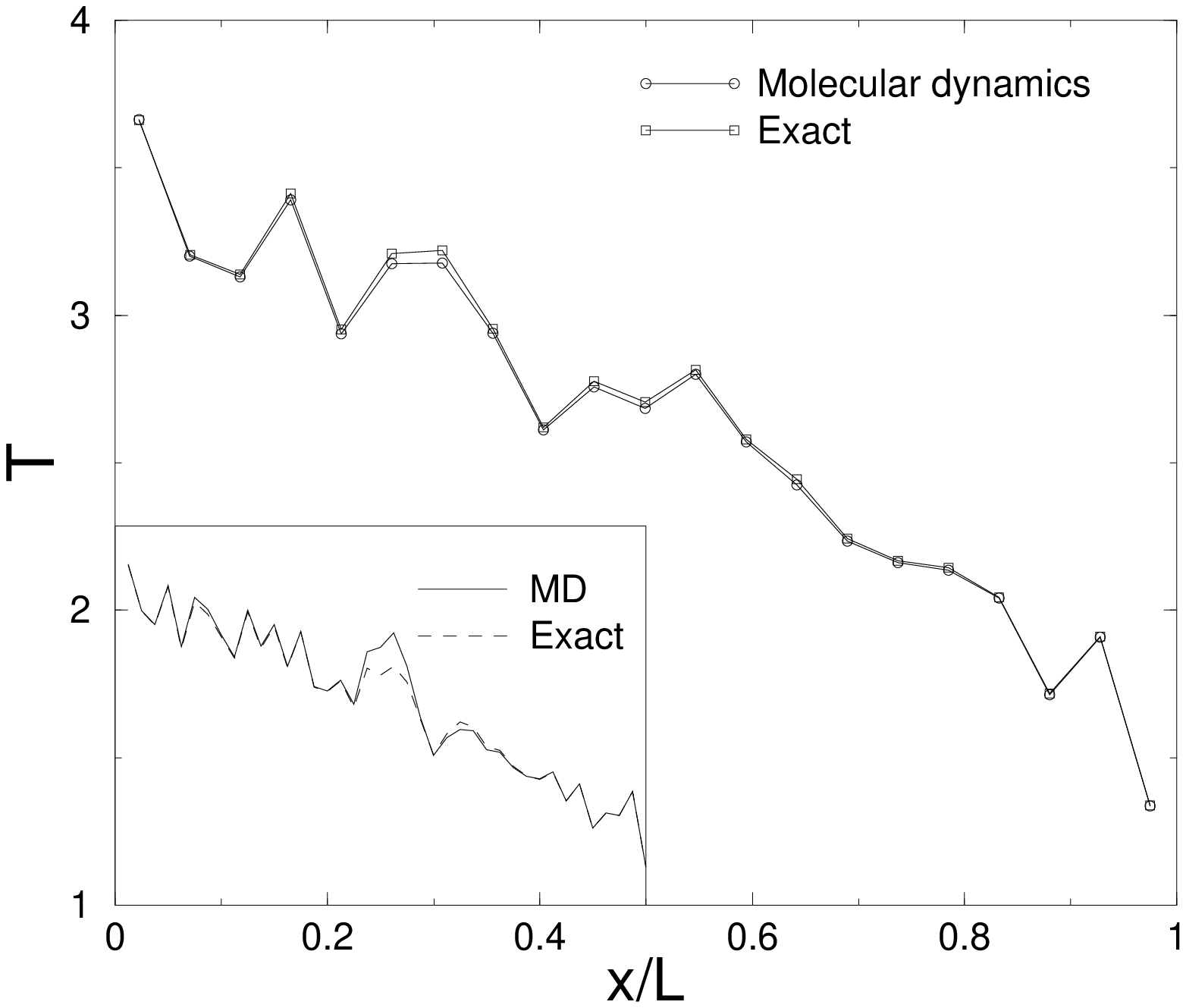}
\begin{figure}
\caption{\label{tprof} The exact temperature profiles as obtained from
inverting Eq.~(\ref{mateq}) are compared with those from molecular
dynamics simulations for two lattice sizes $N=20$ and $N=40$
(inset).  
}
\end{figure}}

\noindent Abhishek Dhar\\
Raman Research Institute\\
Bangalore 560080\\
India.


\begin{references}
\vspace{-1.5cm}
\bibitem{li} B. Li, H. Zhao and B. Hu, Phys. Rev. Lett. {\bf 86}, 63 (2001).
\bibitem{lebo} A. J. O'Connor and J. L. Lebowitz, J. Math. Phys. {\bf 
15}, 692 (1974); A. Casher and J. L. Lebowitz, J. Math. Phys. {\bf 12}, 1701
(1971). 
\bibitem{spohn} H. Spohn and J. L. Lebowitz, Commun. Math. Phys. {\bf
54}, 97 (1977); 
\end{references}
\end{document}